# Anisotropy of the superconducting state parameters and intrinsic pinning in low-level Pr-doped YBa$_2$Cu$_3$O$_{7-\delta}$ single crystals

A Kortyka[1,2], R Puzniak[1], A Wisniewski[1], M Zehetmayer[2], H W Weber[2], Y Q Cai[3], X Yao[3]

[1] Institute of Physics, Polish Academy of Sciences, Aleja Lotników 32/46, PL 02-668 Warsaw, Poland

[2] Atominstitut, Vienna University of Technology, 1020 Vienna, Austria

[3] Department of Physics, Shanghai Jiao Tong University, 800 Dungchuan Road, Shanghai 200240, P. R. China

The influence of low-level Pr substitution in Y$_{1-x}$Pr$_x$Ba$_2$Cu$_3$O$_{7-\delta}$ single crystals on the anisotropy of the superconducting state parameters was investigated by torque magnetometry in magnetic fields of up to 9 T. An averaged anisotropy parameter, $\gamma$, of 7.4 was found and no influence of the Pr ion concentration on $\gamma$ was observed up to a Pr content of 2.4%. A pronounced maximum at angles between 0.5 and 1° out of the $ab$-plane was observed in all crystals in the irreversible angular dependence of the torque. This maximum is attributed to intrinsic pinning associated with kinked vortices. The variation of the irreversible torque with the substitution level indicates the influence of the Pr content on pinning within the CuO$_2$ planes, even though the anisotropy of the superconducting state parameters is not affected.

PACS: 74.25.Bt, 74.25.Ha, 74.72.-h, 75.30.Ha

## I. Introduction

The high-$T_c$ superconductors, HTSC, with layered structure are characterized by a strong anisotropy described by the effective mass anisotropy parameter of the superconducting carriers, $\gamma$, which, in the framework of classical anisotropic Ginzburg-Landau theory, is given by $\gamma = \sqrt{m_c^* / m_{ab}^*} = \lambda_c / \lambda_{ab} = H_{c2}^{ab} / H_{c2}^c = \xi_{ab} / \xi_c$.[1] Here, $m_{ab}^*$ and $m_c^*$ are the effective charge carrier masses related to supercurrents flowing in the $ab$-planes and along the $c$-axis, respectively; $\lambda_{ab}$ and $\lambda_c$ are the corresponding penetration depths, $H_{c2}^{ab}$ and $H_{c2}^c$ the upper critical fields for $H\|ab$-planes and for $H\|c$-axis, and $\xi_{ab}$ and $\xi_c$ the corresponding coherence lengths. In layered HTSC, the coherence length along the $c$-axis is very short, $\xi_c(0) \sim 0.3$ nm, i.e. usually smaller than the distance between adjacent $CuO_2$ layers. Hence, the layer structure itself causes strong pinning of the vortices and intrinsic pinning, which results in a lock-in of vortex lines or kinked state of vortices, appears. The flux lattice penetrating HTSC in the mixed state undergoes dramatic changes when the direction of the magnetic field approaches the $CuO_2$ planes. Four different vortex states are assumed depending on whether the applied field components parallel and perpendicular to the superconducting planes are above or below the respective lower critical fields $H_{c1}^{ab}$ and $H_{c1}^c$ (see ref. 2 and ref. 3 and references therein). The intrinsic pinning mechanism of the vortices comes from a modulation of the order parameter in the direction perpendicular to the layers.[4] Pinning by planar structures occurs when the magnetic field is nearly parallel to the $CuO_2$ planes, namely when the misalignment of the field from the planes does not exceed a critical angle.[5,6] Two critical angles are distinguishable: the first refers to the lock-in of vortex lines ($\theta_{lock}$) and the second to the kinked state of vortices ($\theta_{kink}$). The ordinary Abrikosov-type vortex lattice transforms first into a lattice of kinked Josephson vortices along the $CuO_2$ planes before finally locking in, therefore $\theta_{lock} < \theta_{kink}$.[3] When the pinning by the layered crystalline structure becomes strong enough, i.e. for magnetic fields close to $H\|ab$, the vortices become confined between the $CuO_2$ planes, which causes the lock-in transition. From the viewpoint of thermodynamics, it is more favourable to place the flux lines between superconducting layers than into them. For larger misalignments of the field from the planes, the calculated free-energy density for a tilted vortex lattice[7] coincides with that of the kink model.[8] For a vortex lattice tilted at angles larger than $\theta_{lock}$ from the layer direction, the tilt energy involved by kink formation is balanced by the core energy gained by forming vortex segments parallel to the layers.[9]

Since the YPr123 crystals are twinned, an additional complication may occur. For twinned crystals, the temperature onset of intrinsic pinning occurs above the onset of twin-boundary pinning. Therefore, intrinsic pinning by the layered structure is dominant at temperatures close to the critical temperature, $T_c$, whereas pinning by twin boundaries is dominant at slightly lower temperatures and is present only for vortices aligned with the twin boundaries.[10] Hence, in the present studies performed in the vicinity of $T_c$, pinning by twin boundaries can be neglected. In addition, it is known that the lock-in transition occurs only at fields below $H_{c1}^c$ (see refs. 7,11). The lock-in angle is of the order of $H_{c1}^c/H$ and the transition is difficult to observe even at low fields.[12] Strong vortex pinning reduces the angle of the lock-in transition even further.[13] Taking the conditions of the experiment into account, i.e. the temperature and field range as well as the angular resolution, an observation of the lock-in transition should not be possible. Nevertheless, the vortex core energy modulation remains sizeable even very close to $T_c$ and the total area of the core is $\xi_c \cdot \xi_{ab}$,[12] with $\xi_c > d$ ($d$ denotes the interlayer spacing) and $\xi_{ab} > \lambda_L$ ($\lambda_L$ is the Josephson length, $\lambda_L \approx 5$ nm for Y123). Therefore, intrinsic pinning is not negligible even at high temperatures. Taking the above into account, the only contribution to intrinsic pinning, relevant for the results of this work, is pinning of vortices by the kink structure. An increase of the defect concentration is expected to influence both, the intrinsic pinning due to a change of the parameters $\xi$, $\lambda$, and the thermodynamic critical field $H_c$ (hence the order parameter's modulation), and extrinsic pinning of pancake vortices positioned at defects. It was shown that the critical current density, $J_c$, increases with the defect concentration due to trapping that part of the kink structure by a defect that is perpendicular to the planes (i.e. extrinsic pinning), but saturates at a rather high defect density.[14]

We have shown recently that low level substitution of Pr ions at the Y site in superconducting YBa$_2$Cu$_3$O$_{7-\delta}$ (Y123), with insignificant changes in $T_c$ of the parent compound as a result of such substitutions,[15] introduces effective pinning centres in Y$_{1-x}$Pr$_x$Ba$_2$Cu$_3$O$_{7-\delta}$ (YPr123).[16] On the other hand, it may be expected that introducing additional defects into HTSC, magnetic ions in particular, may be an efficient way to modify the superconducting state anisotropy. For instance, a non-monotonic decrease of the superconducting state anisotropy parameter from $\gamma \approx 6$ to around 3 with increasing Pr-content from 0 to 10% (and simultaneous reduction of $T_c$ by 6 K) was reported.[17] Several groups investigated Pr-substituted Y123, but the influence of low-level substitutions on the anisotropy of the superconducting state parameters and on intrinsic pinning has never been reported. Therefore, investigations of the anisotropy of YPr123 crystals with low level Pr substitutions were investigated.

Only a limited solubility of Pr for Ba has been reported, e.g. at 2.4% Pr only 0.3% of the Pr content substitutes on the Ba site with the remainder substituting on the Y site.[18] Secondly, superconductivity in PrBa$_2$Cu$_3$O$_{7-\delta}$ was reported with $T_c \sim 90$ K for a crystal that did not reach zero resistivity and only 7% of the sample volume showed the Meissner effect.[19] Bulk superconductivity in PrBa$_2$Cu$_3$O$_{7-\delta}$ was reported in single crystals with $T_c$ below 80 K.[20] The results of this work refer to single crystals with $T_c$ and lattice constants comparable to unsubstituted Y123, see below. Therefore, any changes of the intrinsic superconducting parameters due to Pr substitution should to be mainly caused by Pr substitution on Y sites.

In the following, we will report on torque magnetometry measurements of the anisotropy of the superconducting state parameters in low-level Pr-doped YBa$_2$Cu$_3$O$_{7-\delta}$ single crystals. All of the measurements were performed in the vicinity of $T_c$. Therefore, pinning by twin boundaries was neglected. No change in the anisotropy parameter for substitutions of up to 2.4% and no changes in $T_c$ were observed. The influence of increasing the Pr content, the external magnetic field, and the temperature on intrinsic pinning will be presented.

## II. Experimental details

Single crystals of $YBa_2Cu_3O_{7-\delta}$ with small concentrations of Pr up to 2.4% substituted for Y were grown by top seeded solution growth.[21] The crystals were annealed in flowing oxygen at 500 °C for 72 hours. The cooling process was performed either by furnace cooling to room temperature to obtain optimally oxygenated crystals or by slow cooling (20 °C/hour) to obtain crystals with higher oxygen contents. The Pr content was determined by an inductively coupled plasma (ICP) technique. The transition temperature $T_c$ was determined by zero field cooled measurements performed at 1 mT in a 7 T SQUID magnetometer (Quantum Design, MPMS XL). The crystals are plate-like, with $T_c$ ranging from 89.9 K (almost fully oxygenated) to 92.6 K (optimally doped). The XRD analysis (D-5000 Siemens diffractometer using Cu Kα radiation) confirmed the high quality and was used to determine the lattice constants of the crystals. The oxygen content was calculated assuming a linear dependence of the $c$-axis lattice constant on the oxygen content, as proposed by Kruger et al.[22] In the analysis, only the $c$-axis parameter was taken into account due to structural twinning. All crystals with $T_c \approx 90$ K are characterized by very similar lattice constants and, therefore, similar oxygen deficiency. It was already reported that low Pr-contents do not influence the oxygen content in the YPr123 system[23], in agreement with our results. Parameters of the studied crystals are listed in Table I.

The magnetic torque measurements were performed in a 9 T Physical Property Measurement System (Quantum Design, PPMS), equipped with a torque option, in the temperature range $0.94T_c$ – $0.98T_c$ and in magnetic fields of up to 9 T. Experiments at lower temperatures were not possible due to very large torque signals resulting mainly from the irreversible contribution. Measurements performed at temperatures above $T_c$ at various magnetic fields indicated the presence of an anisotropic paramagnetic background leading to a torque with a sinusoidal angular dependence. This magnetic background contribution was taken into account and used to extract the superconducting torque contribution.

## III. Results and discussion

The magnetic torque, $\tau$, was recorded at increasing and decreasing angles $\theta$ between the $c$-axis and the applied magnetic field in an angular range of 180º, in steps of 0.5º. An example of the torque, measured on $Y_{1-x}Pr_xBa_2Cu_3O_{7-\delta}$ with $x = 0.008$ (YPr 0.8%) at $T = 88$ K and $\mu_0H = 2$ T, is presented in Fig. 1. In the angular range 0 – 70º and 110 – 180º, the dependence of the torque for increasing and decreasing angles coincides, while two separate curves are observed in the angular range 70 – 110º. The first angular range corresponds to the reversible torque regime, the second to the irreversible regime. These two regions are distinguishable due to the angular dependence of the irreversibility lines, because $H_{irr}$ in the $H$–$T$ phase diagram is much higher for $H\|ab$ than for $H\|c$, as usual.

The free energy of an anisotropic superconductor in the reversible regime of the mixed state for fields $H_{c1} << H << H_{c2}$ was calculated by Kogan $et$ $al.$[24-26] within the 3D anisotropic London model approach. The corresponding angular dependence of the superconducting torque in the reversible region is given by the first term on the right-hand side of the expression below:

$$\tau(\theta) = -\frac{V\Phi_0 H}{16\pi\lambda_{ab}^2}\left(1-\frac{1}{\gamma^2}\right)\frac{\sin(2\theta)}{\varepsilon(\theta)}\ln\left(\frac{\eta H_{c2}^c}{\varepsilon(\theta)H}\right) + A\sin(2\theta)\cdot \qquad (1)$$

Here, $V$ is the volume of the crystal, $\Phi_0$ is the flux quantum, $\eta$ a numerical parameter of the order of unity depending on the structure of the flux-line lattice and $\varepsilon(\theta)=\left[\cos^2(\theta)+\gamma^{-2}\sin^2(\theta)\right]^{1/2}$. The second term on the right-hand side describes the contribution to the torque from an anisotropic paramagnetic or diamagnetic susceptibility and can be treated as a background contribution to the torque in the superconducting state,[27] with $A$ describing the amplitude of the background torque. By measuring the angular dependence of the torque in the mixed state of a superconductor with anisotropic paramagnetic or diamagnetic background, four parameters can be extracted from the data: the in-plane magnetic penetration depth, the $c$-axis upper critical field, the effective mass anisotropy, and the background torque amplitude.

Note that the first term on the right-hand side of Eq. (1) describes the reversible torque and that Eq. (1) is generally valid only in the magnetic field range $H_{c1} << H << H_{c2}$. In order to obtain fully reversible curves, the measurements should be made in magnetic fields above the irreversibility field in the $ab$-plane, $H_{irr}^{ab}$, but much below $H_{c2}^c$. For highly anisotropic superconductors with an irreversibility line at high magnetic fields, it is often difficult to fulfil these conditions in the full angular range. In such cases, $\tau_{rev}(\theta)=\left(\tau(\theta^+)+\tau(\theta^-)\right)/2$ is calculated from data obtained by clockwise and counterclockwise rotating the sample in the magnetic field, see Fig. 1. In the case of a pronounced irreversibility of the clockwise and counterclockwise torque, this procedure may lead to an overestimation of the anisotropy parameter, see ref. 28 and references therein.

Thermal fluctuations of vortices are not considered directly in Eq. (1). As already shown, Schneider's functional[29] for a fluctuating torque is estimated to be sinusoidal close to $T_c$. Therefore, it has the same dependence as the anisotropic paramagnetic/diamagnetic background contribution and can be described by the second term on the right-hand side of Eq. (1). The temperature of the 3D-2D crossover can be estimated by applying the relation $T_{2D-3D} \approx T_c$ [$1-2\xi_{ab}^2/(\gamma d)^2$] (see ref. 30) and occurs in the temperature range 61 – 81 K in our crystals. The parameters used in the calculation are as follows: $\gamma = 5 - 8$, $\xi_{ab} = 1.6$ nm (see ref. 20), $T_c = 89.9 - 92.6$ K, $d = 0.8$ nm.[7] Since the spatial period of the intrinsic pinning potential coincides with the crystal lattice constant, there is just a single minimum energy well for a vortex within the unit cell in the middle of the bilayer spacing.[31] The crossover field corresponding to the 3D-2D crossover is given by $H_{cr} = \Phi_0/(\gamma d)^2$ (see ref. 1), and is in the range 50 – 130 T, i.e. at values much higher than those reached in our experiments. As a result, the assumption of continuous vortex lines and cores, made in the 3D anisotropic model (Eq. (1)), is valid in the full range of temperatures and magnetic fields of our measurements.

The most reliable parameter which can be extracted from the torque data is the anisotropy parameter. The upper critical field is difficult to extract since it enters only logarithmically into the Eq. (1). Additionally, the temperature and field dependence of a pronounced angular dependent background may strongly affect the extracted $H_{c2}^{c}$ values. Therefore, $H_{c2}^{c}$ was fixed in the fitting procedure to the $H_{c2}^{c}(T)$ value obtained from magnetization scaling[16] performed on the same crystals (before cutting them to a smaller size) by SQUID measurements[16], see Table II. In the present work, we set $\eta = 1$. It is well known that $\eta$ should be of the order of unity, but setting $\eta = 1$ is in some sense arbitrary. Nevertheless, it does not affect the estimated anisotropy too much, since the anisotropy is quite insensitive to variations of $H_{c2}$.[32] Three other parameters were extracted simultaneously from a fit of Eq. (1) to the averaged angular torque dependence. The temperature dependence of the penetration depth extracted by fitting Eq. (1) to the experimental data agrees reasonably well with that obtained from magnetization scaling ($\lambda_{ab}$ at $T = 0.97T_c$: for YPr 1.3%: $4.81 \cdot 10^{-7}$ m (torque), $4.26 \cdot 10^{-7}$ m (scaling); for 90K-Y123: $5.64 \cdot 10^{-7}$ m (torque), $4.55 \cdot 10^{-7}$ m (scaling)). Nevertheless, since magnetization scaling was performed in a much wider temperature range and on larger crystals, more accurate values for the penetration depth are expected to result from magnetization scaling[16] than from fitting the torque data. The upper critical field parameters and the zero temperature penetration depths obtained from magnetization scaling[16] are listed in Table II. The values of both $H_{c2}^{c}$ and $\lambda_{ab}$ for the unsubstituted crystals agree very well with the data obtained by other authors.[33,34] The temperature dependence of $A(T)$, describing the background amplitude of an anisotropic normal state contribution to the torque, estimated by fitting the $A$ value at different temperatures, is quite well approximated by $A(T) = C_1 / T + C_2$, which is consistent with Curie-Weiss paramagnetism.

In the irreversible torque recorded for both branches of the angular dependence, under clockwise and counterclockwise rotating the sample in the magnetic field (Fig. 1), two well separated maxima are visible. The first is relatively wide and located quite far from the $ab$ plane. Its angular position is mainly related to the anisotropy parameter. The reversible torque in the angular range corresponding to the first torque maximum is well described by Eq. (1). The second maximum is located very close to the $ab$ plane. The reversible torque at angles very close to 90° cannot be described within the 3D model and the second maximum in the torque dependence is attributed to intrinsic pinning (see below).

The angular dependence of the reversible torque, normalized to the first torque maximum obtained for two Y-123 samples with different oxygen content at almost the same reduced temperature $T/T_c = 0.97$ and in the same external magnetic field, is presented in Fig. 2. Both curves overlap almost perfectly, i.e. the extracted anisotropy parameters of both crystals, one optimally doped with $\delta = 0.085$ and one overdoped with $\delta = 0.027$, are very similar, see the inset of Fig. 2. On the other hand, it is known that the anisotropy decreases at higher oxygen content in Y123.[36] Such dependence was not found in the present crystals, probably due to the fact that the oxygen content in both crystals was close to the optimal oxygenation level. Despite the very weak, if any, influence of the oxygen content on the anisotropy (at small oxygen deficiency), a comparison of the anisotropy parameters in the Pr substituted crystals was made between crystals with approximately the same oxygen deficiency. For the crystals with $\delta \approx 0.03$, no change of the anisotropy parameter was found within experimental accuracy for Pr-contents between 0 and 2.4%, see Table II and Fig. 3. The anisotropy dependence on the Pr content for all crystals with $\delta$ in the range between 0.027 and 0.085 is presented in the inset of Fig. 3. Examples of the torque data obtained at different temperatures for 92K-Y123 are presented in Fig. 4 and examples of the torque data in different fields for YPr 0.8% in Fig. 5. The insets of these Figures present the temperature dependence of the anisotropy parameter for 92K-Y123 and the field dependence of the anisotropy parameter for YPr 0.8%. Here, the error of the anisotropy was estimated from the difference between the anisotropy parameter derived from the fit of Eq.(1) to the reversible, i.e. averaged, torque and the parameter derived from the fit of Eq.(1) to the clockwise or counterclockwise angular dependence of the torque only. Thus, the quoted error represents the upper limit of errors due to the torque hysteresis, which is pronounced at low temperatures (see inset of Fig. 4) and at low fields (see inset of Fig. 5). Neither a temperature nor a field dependence of the anisotropy was found within experimental accuracy. This is consistent with the absence of a temperature or a field dependence of the anisotropy parameter in the cuprate superconductors, as expected for single gap superconductors, and in contrast to $MgB_2$ (ref. 37) and the pnictides, where two superconducting energy gaps were proposed.[38] However, it should be noted that, due to significant pinning, our measurements were performed in a relatively narrow temperature range, much narrower than that accessible for $MgB_2$ and the pnictides and, therefore, the data cannot completely rule out a temperature or field dependence of the anisotropy parameter.

Finally, the superconducting state anisotropy parameters averaged over all temperatures and fields investigated for each crystal together with the upper critical field parameters and the zero temperature penetration depths obtained from magnetization scaling[16] are presented in Table II.

A consequence of the layered structure of the cuprates is the intrinsic pinning of the vortex lines between the $CuO_2$ planes observed as a large peak in the torque signal close to the $ab$ plane (which we will call the kinked structure angle $\theta_{kink} \approx 90°$ in contrast to the lock-in angle), see Fig. 6. The experimental verification of intrinsic pinning remains difficult, see e.g. ref. 39. It is known that weak intrinsic pinning exists in optimally doped Y123 single crystals.[40] In order to estimate the effect of Pr ions on pinning close to the $ab$-plane geometry, the torque was compared for two crystals with the same oxygen deficiency. Figure 7 presents the angular dependence of the torque observed for YPr 2.4% and 90K-Y123 at different external magnetic fields and temperatures.

Kugel $at$ $al.$[41] analyzed the structure of the pinned flux line lattice (FLL) in terms of competing pinning mechanisms, i.e. bulk pinning (3D) by point defects and intrinsic pinning (2D) within the $ab$-plane. A non-trivial coupling between 2D and 3D pinning, depending on the relation between the Gibbs free energy per unit volume due to bulk ($G_{3D}$) and intrinsic pinning ($G_{2D}$), was found.[41] The FLL was proposed to have two stable configurations, one corresponding to 3D pinning (i.e. $|G_{3D}| > |G_{2D}|$) and one for intrinsic or 2D pinning (i.e. $|G_{3D}| < |G_{2D}|$). Changes in magnetic field, temperature and material parameters can lead to a crossover between these configurations.

The torque maximum corresponding to vortex pinning by the kinked structure was very pronounced for all crystals already at temperatures very close to $T_c$, which is due to strong intrinsic pinning. It was found that the angle $\theta_{kink}$ varies between 90.5 and 91°, depending on crystal, and decreases at higher temperatures. With increasing magnetic ion concentration, an increase of both intrinsic and bulk pinning is expected, the latter was confirmed in our previous work.[15] Presumably, pinning at angles $\theta \leq \theta_{kink}$ should be associated with 2D (intrinsic) pinning, whereas with 3D (bulk) pinning at angles $\theta > \theta_{kink}$. Analyzing the data of Fig. 7, we note a larger irreversibility of the torque at lower temperature for the Pr-doped crystal than for the pure Y123 crystal. At the same time the angular range, where vortex pinning is efficient for YPr 2.4% is narrower than that for 90K-Y123. Intrinsic pinning seems to be stronger in 90K-Y123 than in YPr 2.4% since the maximum, corresponding to vortex pinning by kinks, diminish faster with increasing field (i.e. vortex lattice changes from kinked to tilted) in Pr-doped crystal as in pure Y123, where kinks can be traced up to higher fields. This is probably related to a higher rate of nucleation of kinks at the defect sites as a result of Pr substitution in YPr 2.4%. The part of the kink structure parallel to the $c$-axis, pinned by defects located in the planes, may be the reason for the previously observed[15] increase of the critical current at higher Pr concentration. Then, the critical current density due to trapping of perpendicular to the planes part of the kink structure saturates faster (see ref. 14) for YPr2.4% than for pure Y123. This is probably related to the observed[16] shift of the magnetic field, at which the magnetization maximum corresponding to the fishtail effect appears, to lower fields for higher Pr content in YPr123.

## IV. Conclusions

The superconducting state anisotropy parameter is found to be independent of the Pr content in the investigated low-level Pr-doped YBa$_2$Cu$_3$O$_{7-\delta}$ crystals with Pr contents of up to 2.4%. The averaged value of the anisotropy is 7.4 and independent of temperature and field. Intrinsic pinning was observed in all crystals via a pronounced peak of the irreversible torque signal, when the magnetic field deviates by $0.5 - 1°$ from the direction parallel to the $ab$-plane. The peak in the torque was attributed to the kinked state of vortices. The intrinsic pinning was found to be stronger in pure Y123 than in YPr 2.4%, probably due to a higher nucleation rate of kinks at the defects in the Pr-doped crystal.

## Acknowledgements

We would like to thank V. Domukhovski for the x-ray measurements. AK thanks the European NESPA project for financial support. This work was partially supported by the Polish Ministry of Science and Higher Education under the research projects No. N N202 4132 33 and No. N N202 2412 37. X. Y. thanks the Shanghai Committee of Science and Technology Grants.

**Table I.** Superconducting $T_c$, lattice constants, oxygen deficiency, and dimensions of the investigated crystals.

| Sample | Symbol | $T_c$ (K) | $a, b, c$ (nm) | $\delta$ | $x \times y \times z$ (mm$^3$) |
|---|---|---|---|---|---|
| YBa$_2$Cu$_3$O$_{7-\delta}$[a] | 92K-Y123 | 92.6 | 0.38197, 0.38886, 1.16925 | 0.085[c] | $1.05 \times 1.00 \times 0.40$ |
| YBa$_2$Cu$_3$O$_{7-\delta}$[b] | 90K-Y123 | 90.1 | 0.38178, 0.38857, 1.16840 | 0.027[c] | $1.13 \times 1.09 \times 0.35$ |
| Y$_{0.992}$Pr$_{0.008}$Ba$_2$Cu$_3$O$_{7-\delta}$[a] | YPr 0.8% | 91.2 | 0.38188, 0.38869, 1.16901 | 0.066[c] | $1.11 \times 1.05 \times 0.55$ |
| Y$_{0.987}$Pr$_{0.013}$Ba$_2$Cu$_3$O$_{7-\delta}$[b] | YPr 1.3% | 89.9 | 0.38173, 0.38858, 1.16846 | 0.029[c] | $1.03 \times 0.98 \times 0.35$ |
| Y$_{0.976}$Pr$_{0.024}$Ba$_2$Cu$_3$O$_{7-\delta}$[b] | YPr 2.4% | 90.2 | 0.38177, 0.38867, 1.16849 | 0.027[c] | $1.15 \times 0.93 \times 0.30$ |

[a] Furnace cooling to room temperature
[b] Slow (i.e. 20 °C/hour) cooling applied to obtain the lowest oxygen deficiency
[c] Calculated from ref. 22

**Table II.** Superconducting $T_c$, oxygen deficiency, upper critical field parameters, zero temperature penetration depth, and average superconducting state anisotropy parameter (with statistical errors) for the investigated crystals.

| Sample | $T_c$ (K) | $\delta$ | $\mu_0 dH_{c2}^c/dT\vert_{Tc}$ (T/K)[a] | $\mu_0 H_{c2}^c(0)$ (T)[b] | $\lambda_{ab}(0)$ (nm)[a] | $\gamma$ |
|---|---|---|---|---|---|---|
| 92K-Y123 | 92.6 | 0.085 | - 2.00 | 134.4 | 96.2 | 7.25(0.29) |
| 90K-Y123 | 90.1 | 0.027 | - 1.83 | 119.6 | 86.3 | 7.17(0.17) |
| YPr 0.8% | 91.2 | 0.066 | - 1.93 | 127.7 | 105.4 | 7.87(0.21) |
| YPr 1.3% | 89.9 | 0.029 | - 2.00 | 130.5 | 84.3 | 7.38(0.30) |
| YPr 2.4% | 90.2 | 0.027 | - 2.09 | 136.8 | 84.0 | 7.38(0.47) |

[a] from magnetization scaling, see ref. 16
[b] assuming clean limit and WHH dependence[35]

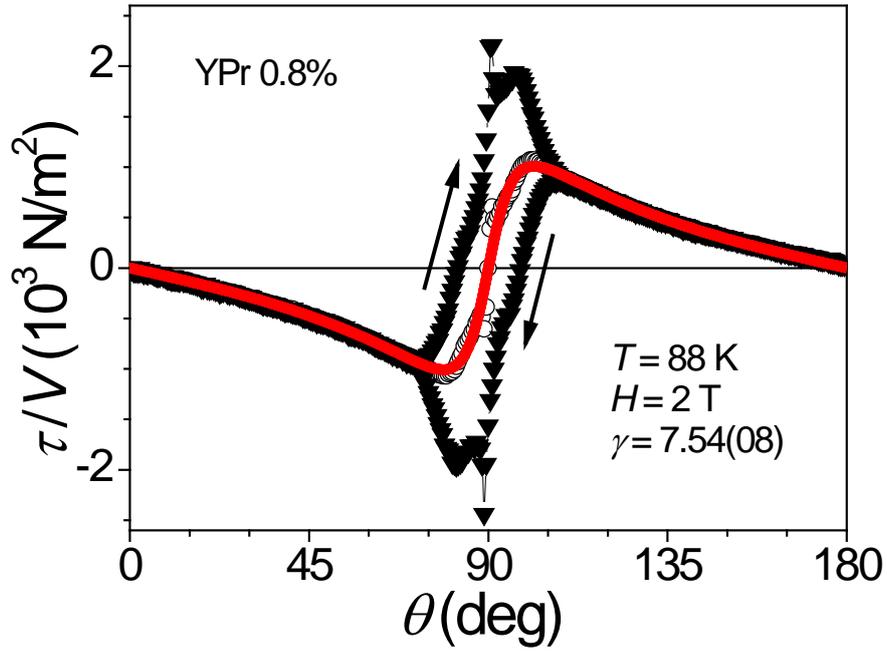

Fig. 1. Angular dependence of the torque for Y123 with 0.8% Pr content. Triangles denote the irreversible (clockwise and counterclockwise) and circles denote the reversible (averaged) torque. The solid line shows a fit of equation (1) to the data.

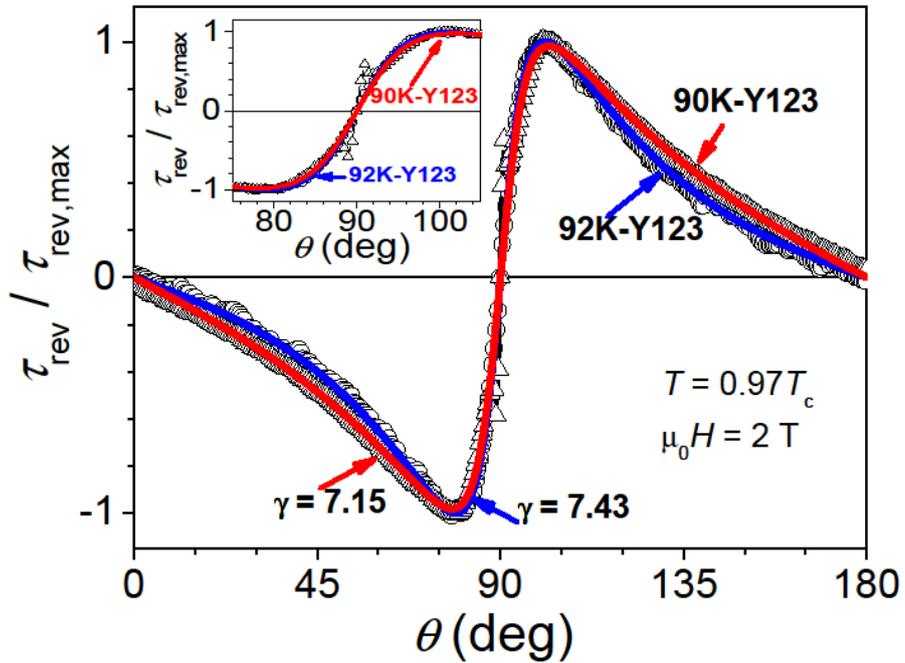

Fig. 2. Angular dependence of the normalized reversible torque for 90K-Y123 (triangles) and 92K-Y123 (circles) with fits of equation (1) to the data. Inset: the same dependence for angles close to 90 deg.

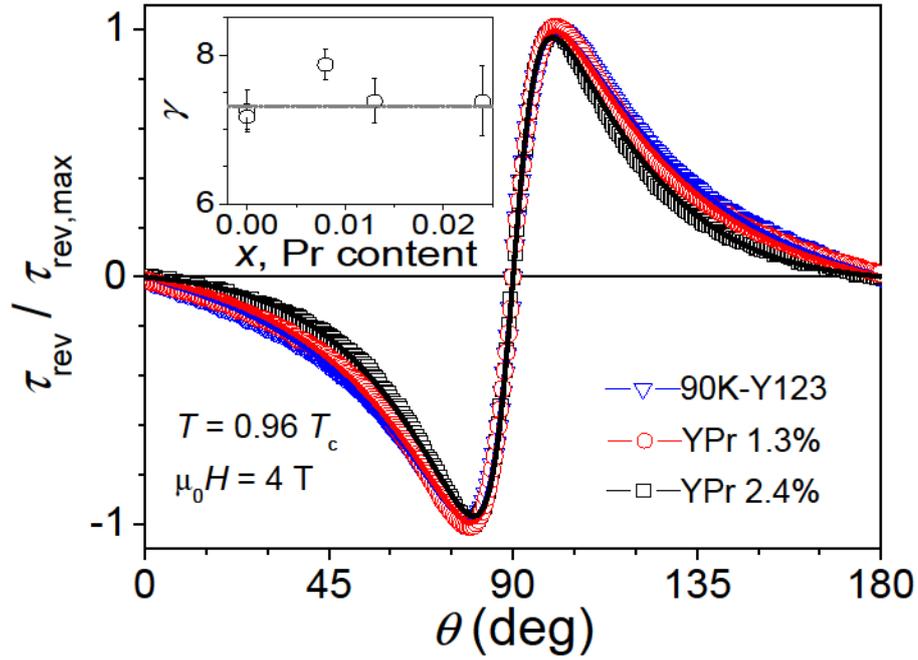

Fig. 3. Angular dependence of the normalized reversible torque and the corresponding fit of equation (1) to the data: 90K-Y123 (triangles), YPr 1.3% (circles), YPr 2.4% (squares). Inset: Dependence of the anisotropy on the Pr content for all crystals, the line corresponds to the averaged anisotropy parameters for crystals with $\delta \approx 0.03$.

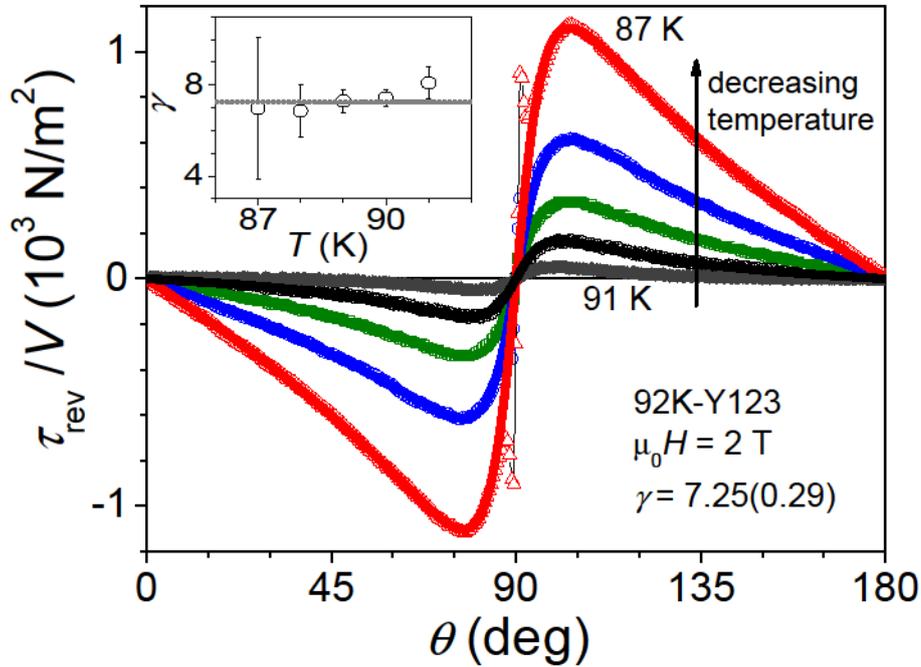

Fig. 4. Angular dependence of the reversible torque for 92K-Y123 at temperatures from 87 to 91 K at $\mu_0 H = 2$ T and the corresponding fit of equation (1) to the data. Inset: Temperature dependence of the anisotropy parameter with maximal errors due to torque hysteresis. The line is a guide to the eye.

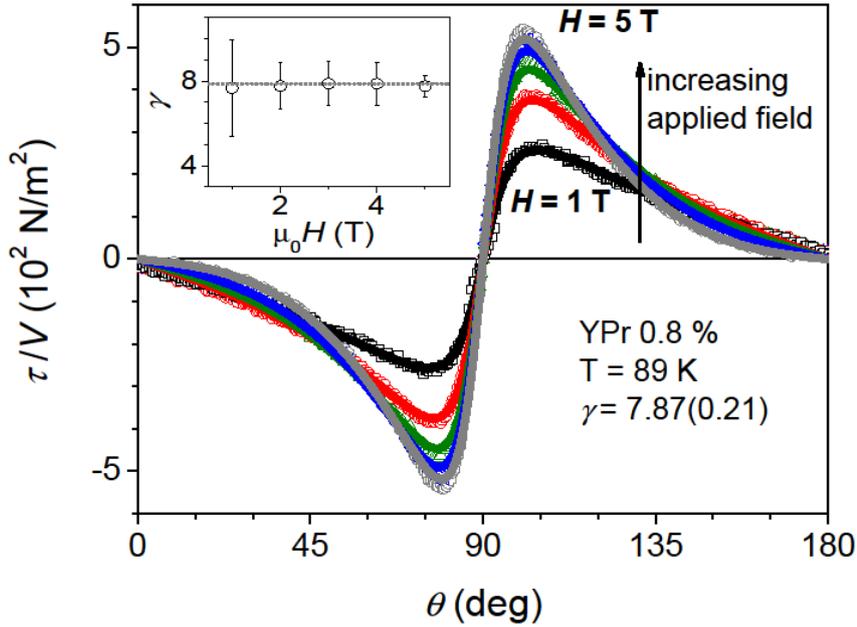

Fig. 5. Angular dependence of the reversible torque for YPr 0.8% at applied magnetic fields between 1 and 5 T at $T = 89$ K and the corresponding fit of equation (1) to the data. Inset: Field dependence of the anisotropy parameter with maximal errors due to torque hysteresis. The line is a guide to the eye.

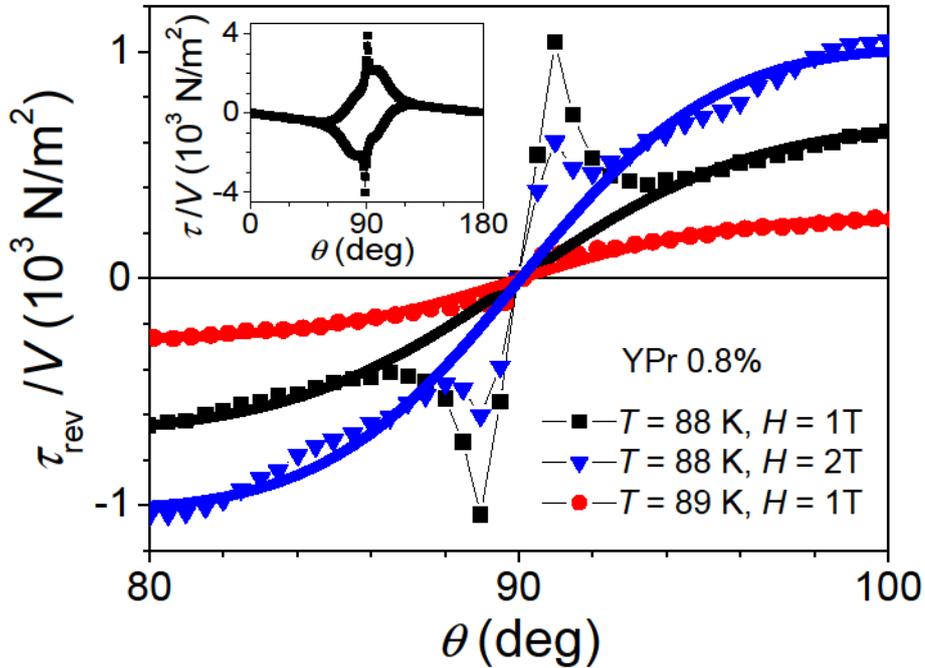

Fig. 6. Angular dependence of the reversible torque for YPr 0.8% at two different temperatures and fields, obtained by averaging the irreversible torque dependence upon clockwise and counterclockwise rotating the sample, indicating deviations from the dependence according to Eq. (1) (represented by straight lines). Inset: Example of the angular dependence of the irreversible torque with pronounced intrinsic pinning close to the *ab*-plane.

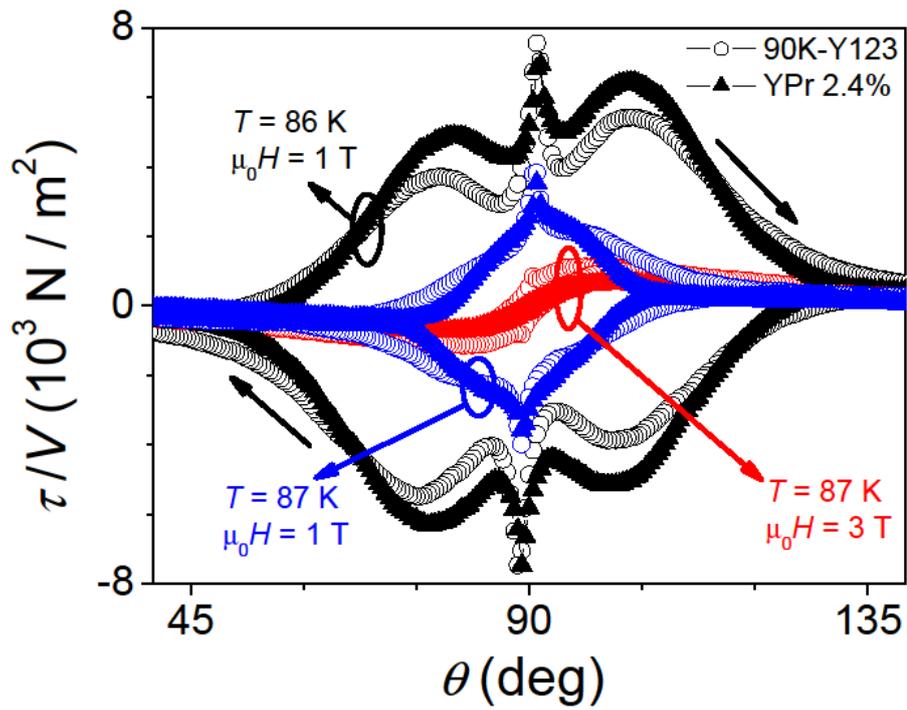

Fig. 7. Angular dependence of the irreversible torque for two crystals with the same oxygen content; open symbols: 90K-Y123, full symbols: YPr 2.4%. The torque maxima corresponding to intrinsic pinning at $\theta^* \approx 90°$ are clearly visible.